\documentclass[twocolumn,showpacs,preprintnumbers,amsmath,amssymb]{revtex4}
\usepackage{graphicx}
\usepackage{dcolumn}
\usepackage{bm}

\begin{document}


\title{The ground state of a mixture of two species of fermionic atoms
in 1D optical lattice}

\author{Shi-Jian Gu} \author{Rui Fan} \author{Hai-Qing Lin}
\affiliation{Department of Physics and Institute of Theoretical
Physics, The Chinese University of Hong Kong, Hong Kong, China}

\begin{abstract}
In this paper, we investigate the ground state properties of a mixture of two
species of fermionic atoms in one-dimensional optical lattice, as described by
the asymmetric Hubbard model. The quantum phase transition from density wave to
phase separation is investigated by studying both the corresponding charge
order parameter and quantum entanglement. A rigorous proof that even for the
single hole doping case, the density wave is unstable to the phase separation
in the infinite $U$ limit, is given. Therefore, our results are quite
instructive for both on-going experiments on strongly correlated cold-atomic
systems and traditional heavy fermion systems.
\end{abstract}


\date{\today}
\maketitle

\section{Introduction}
Rapid progress in Bose-Einstein condensates in optical lattices
\cite{Bloch,MHolland01,MGreiner02,OMandel03,CChin04} has opened fascinating
experimental possibilities in condensed matter physics, atomic physics, and
quantum information. For example, the experiment on neutral atoms trapped in
the periodic potential of an optical lattice has been used to realize an array
of quantum gates \cite{OMandel03}. Moreover, cold-atomic systems are
intrinsically related to many-body models in condensed matter physics. Compared
to solid state systems, cold-atomic systems could be better experimentally
controllable. Thus, the investigation of cold-atomic systems will not only help
us to have a deep understanding of known physical phenomena in many-body
systems, but also provide hints to explore new areas of physics. Such a
beautiful prospect has attracted many theoretical attentions
\cite{DJaksch98,Hofstetter02,Liu04,MACazalilla05,LMDuan05,RBDiener06}.

Quite recently, experiments on fermionic atoms trapped in optical
lattices\cite{CChin04} were carried out which opened a door for us to find
deeper insights into some essential problems in condensed matter physics, such
as BEC-BCS crossover, superfluidity, and Mott-insulating phase. It was proposed
that ultracold fermionic atoms exposed to the periodical potential of an
optical lattice could be an ideal realization of the Bose-Hubbard model
\cite{DJaksch98}, the spin-dependent Hubbard model \cite{Liu04}, and the
antiferromagnetic states or $d$-wave pairing states \cite{Hofstetter02}. The
unique control over all relevant parameters in these systems \cite{CChin04}
allows people to carry out experiments which are not handy with solid state
systems, so they marked a milestone towards the understanding of some
fundamental concepts in quantum many-body systems.

In this paper, we consider a system of two species of fermionic atoms
\cite{MACazalilla05} with equal numbers (or one type of fermionic atoms with
spin-depent hoping integral\cite{Liu04}) away from half-filling in an
one-dimensional optical lattice, as described by the asymmetric Hubbard model.
\cite{UBrandt91,GFath95,CAMacedo02} (AHM). The system is expected to have a
density wave (DW) state and phase separation (PS) of two atom species state
\cite{VJEmery90}, and we investigate the quantum phase transition (QPT) from
the DW state to the PS state in this system by studying both the quantum
entanglement and traditional DW order parameter. We show that the entanglement
can help us to witness critical phenomenon and shows scaling behavior around
the critical point. The phase transition is also clarified by the competition
between two different modes of structure factor. A global phase diagram as a
function of the local interaction $U$ and the ratio of two hoping integrals is
then obtained under different filling conditions. Moreover, we give a rigorous
proof, that even for the case of a single hole doping away from half-filling,
the DW state is unstable to the PS state in the infinite $U$ limit. As will be
shown below, if we regard two regions in the PS phase as one solid-like region
of heavy atoms and another as a liquid-like region of light atoms,
respectively, the QPT is just a physical realization of the quantum solvation
process \cite{FPaesani01} in the optical lattice. Therefore, our results are
quite instructive for on-going experiments on strongly correlated cold-atomic
systems. The behavior of entanglement in this system can help people to have a
deep understanding of the critical phenomenon.

This paper is organized as follows. In section \ref{sec:model}, we introduce
the Hamiltonian of the AHM, and show how to realize the model in the quasi
one-dimensional periodical potential of an optical lattice. We will also
briefly introduce the background of the model in the condensed matter physics.
In section \ref{sec:ent}, we study the ground-state entanglement of the system.
We will show that a schematic phase diagram can be obtained from the
entanglement between a local part and the rest of the system of a finite
sample. In section \ref{sec:cdwo}, by studying the structure factor of the
density distribution of heavy atoms, we can obtain a quantitative phase diagram
for different filling conditions via both the exact diagonalization (ED) and
density matrix renormalization group (DMRG) methods. In section
\ref{sec:onehole}, we will give a rigorous proof that even for the case of a
single hole doping away from half-filling, the DW state is unstable to the PS
state in the infinite $U$ limit. While if $U$ is very large, the critical point
then is approached linearly with $1/U$. In section \ref{sec:discuss}, we will
discuss the mechanism of the existence of the PS, the possibility of the PS in
high dimension, and conditions for experimental realization. Finally, we
summarize our results in section \ref{sec:sum}.

\section{The model Hamiltonian}
\label{sec:model}

The one-dimensional AHM is defined as
\begin{eqnarray}\label{eq:Hamiltonian}
H=-\sum_{j=1}^{L}\sum_{\delta=\pm 1} \sum_\sigma t_\sigma
c^\dagger_{j,\sigma}c_{j+\delta, \sigma}+U \sum_{j=1}^L n_{j, \alpha}n_{j,
\beta}.
\end{eqnarray}
In Eq.(\ref{eq:Hamiltonian}), $t_\sigma$ ($\sigma=\alpha, \beta$) distinguishes
the species of fermionic atoms (e.g., $^6$Li and $^{40}$K),
$c^\dagger_{j,\sigma}$ and $c_{j,\sigma},\sigma=\uparrow,\downarrow$ are
creation and annihilation operators for $\sigma$ atoms at site $j$
respectively, and $n_\sigma=c_\sigma^\dagger c_\sigma$, while $U$ denotes the
strength of on-site interaction. In this model, the Hamiltonian has
U(1)$\otimes$U(1) symmetry for the general $t_\sigma$, and the atoms number
$N_\alpha=\sum_j n_{j,\alpha}, N_\beta=\sum_j n_{j, \beta}$ are conserved
respectively. The total number of atoms is given by $N=N_\alpha+N_\beta$, and
the filling factor is $n=N/L$.

The asymmetric Hubbard model (\ref{eq:Hamiltonian}) can be used as an effective
model to describe a mixture of two species of fermionic atoms in an
one-dimensional optical lattice. In order to have a quasi one-dimensional
system, we suggest that the optical lattice potential takes the form of
\begin{eqnarray}
&&V(x, y, z)=V_0\sin^2 (k x) + V_\perp[\sin^2(k y) + \sin^2 (k z)],\nonumber \\
&&V_0=\nu\frac{\hbar^2 k^2}{2m},\nonumber \\
&&V_\perp=\nu_\perp \frac{\hbar^2 k^2}{2m}.
\end{eqnarray}
Here $k=2\pi/\lambda$ and $\lambda$ is the wavelength of the laser light, and
$V_0$ and $V_\perp$ denote the maximum potential depth along the $x$ direction
and in the $yz$ plane respectively. The potential depth is measured in units of
the recoil energy ${\hbar^2 k^2}/{2m}$. In order to freeze the hoping process
in the $yz$ plane we should have $V_\perp \gg V_0$. For a single atom in the
periodic lattice, its wave function is the Bloch state, which is actually a
superposition of well localized Wannier state. Therefore, if we restrict
ourself to a very low temperature, where the thermal fluctuation cannot excite
the atom to the second band, the Wannier state can be approximated by the
ground state of a single atom in the potential well. For the present case, the
ground state can be written as
\begin{eqnarray}
&&\Psi_0(x, y, z)\simeq \left(\frac{m\omega_\perp}{\pi\hbar}\right)^{1/2}
e^{-\frac{m\omega_\perp}{2\hbar}(y^2+z^2)} \Psi_0(x), \label{eq:groundstate1}
\end{eqnarray}
where
\begin{eqnarray}
&&\Psi_0(x)=\left(\frac{m\omega}{\pi\hbar}\right)^{1/4}
e^{-\frac{m\omega}{2\hbar}x^2}\nonumber \\
&&\omega_\perp=\frac{\hbar k^2\sqrt{\nu_\perp}}{m},\;\;\; \omega=\frac{\hbar
k^2\sqrt{\nu}}{m}.
\end{eqnarray}
Then the hoping matrix element between the two adjacent sites $i$, $j$ can be
calculated as
\begin{eqnarray}
t=-\int d{\bf r}\, w_\sigma({\bf r}-{\bf r}_i)\left(-\frac{\hbar^2 \nabla^2}{2m}
+V\right)w_\sigma({\bf r}-{\bf r}_{j}),
\end{eqnarray}
which results in the hoping integral along the $x$ direction
\begin{eqnarray}
t\simeq \frac{\hbar^2 k^2}{2m}(\sqrt{v}+2\sqrt{v_\perp})e^{-\pi^2\sqrt{v}}.
\end{eqnarray}

Moreover, if two atoms, $\alpha$ and $\beta$ occupy the same site, they will
repel each other. The on-site interaction can be approximated with
\begin{eqnarray}
U\simeq\frac{4\pi\hbar^2 a}{\sqrt{m_\alpha m_\beta}}\int |w_\alpha({\bf r})|^2
|w_\beta({\bf r})|^2 d{\bf r}\nonumber,
\end{eqnarray}
where $a$ is the scattering length. Using the wave function of Eq.
(\ref{eq:groundstate1}), we obtain
\begin{eqnarray}
U\simeq \frac{4\pi\hbar^2 a}{\sqrt{m_\alpha m_\beta}} \frac{k
v^{1/4}}{\sqrt{\pi}} \frac{k^2 v_\perp^{1/2}}{\pi}.
\end{eqnarray}
Finally, if we have a system of two species of polarized fermionic atoms in the
optical lattice, the hoping integral and the on-site interaction will have the
form (in units of $t_\alpha$)
\begin{eqnarray}
&&\frac{t_\beta}{t_\alpha}=\frac{m_\alpha}{m_\beta},\nonumber \\
&&\frac{U}{t_\alpha}=\frac{16 a\sqrt{\pi m_\alpha/m_\beta}}{\lambda}
\frac{v^{1/4}
v_\perp^{1/2}}{(\sqrt{v}+2\sqrt{v_\perp})}\;e^{\pi^2\sqrt{v}}.
\label{eq:tUrelation4}
\end{eqnarray}
Taking Li($\alpha$) and K ($\beta$), two species of atoms, as an example and
$v_\perp=16$, we have
\begin{eqnarray}
&&{t_\beta}\simeq 0.15,\nonumber \\
&&{U}\simeq \frac{35.87\; a}{\lambda} \frac{v^{1/4}
}{(\sqrt{v}+8)}\;e^{\pi^2\sqrt{v}}.
\end{eqnarray}

In condensed matter physics, the asymmetric Hubbard model is one of the most
simplest two band models which is believed to describe many essential physical
properties of strongly correlated systems. To understand the interesting
phenomena which may happen in the ground state of the Hamiltonian
(\ref{eq:Hamiltonian}), it is very useful to look into the two limiting cases
of Eq. (\ref{eq:Hamiltonian}). If $t_\alpha=t_\beta$, the AHM becomes the
Hubbard model \cite{Hubbard}. In 1D, the Hubbard model can be solved exactly by
the Bethe-ansatz method.\cite{ELieb} The wave function and the energy spectra
then can be calculated exactly. In the large $U$ limit, the Hubbard model can
be approximated by the famous $t-J$ model, in which the spin-spin interaction
is of the antiferromagnetic type. Therefore, it is widely accepted that the
ground state of the Hubbard model at half-filling shows the spin-density wave.
On the other hand, if $t_\beta=0$, the AHM becomes Falicov-Kimball model.
\cite{LMFalicov69,PLemberger92,JKFreerichs03,TKennedy86} In 1D, it has been
pointed out that the system will segregate into an empty lattice (with no
$\beta$ atoms and all $\alpha$ atoms) and a full lattice (with all $\beta$
atoms and no $\alpha$ atoms) in the large $U$ limit when away from the
half-filling. Therefore, the two limiting cases of the AHM belong to different
universality classes, a phase transition from PS to DW is expected to appear
somewhere on the $U-t_\beta$ plane.

\begin{figure*}
\includegraphics[width=7.5cm]{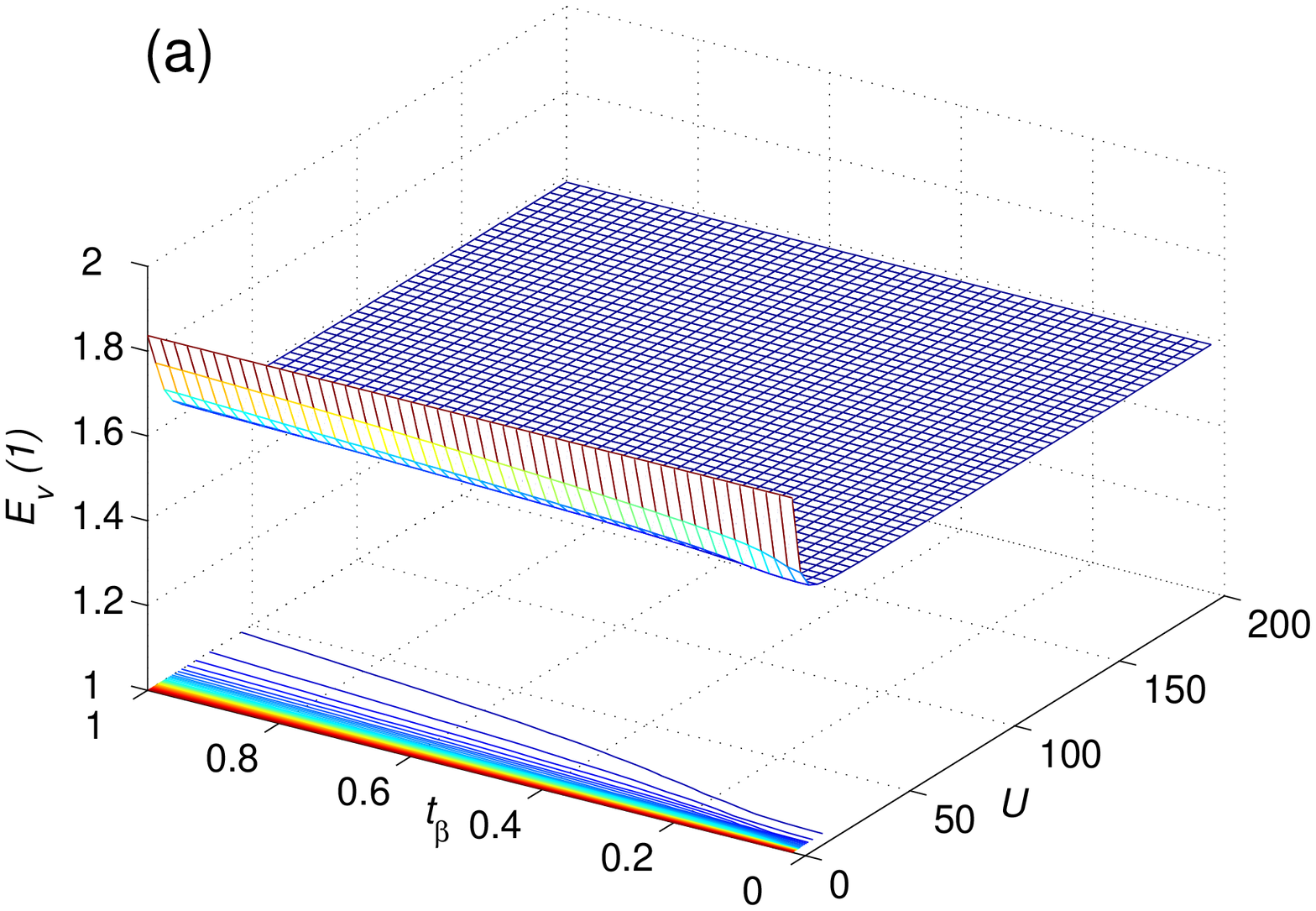}
\includegraphics[width=7.5cm]{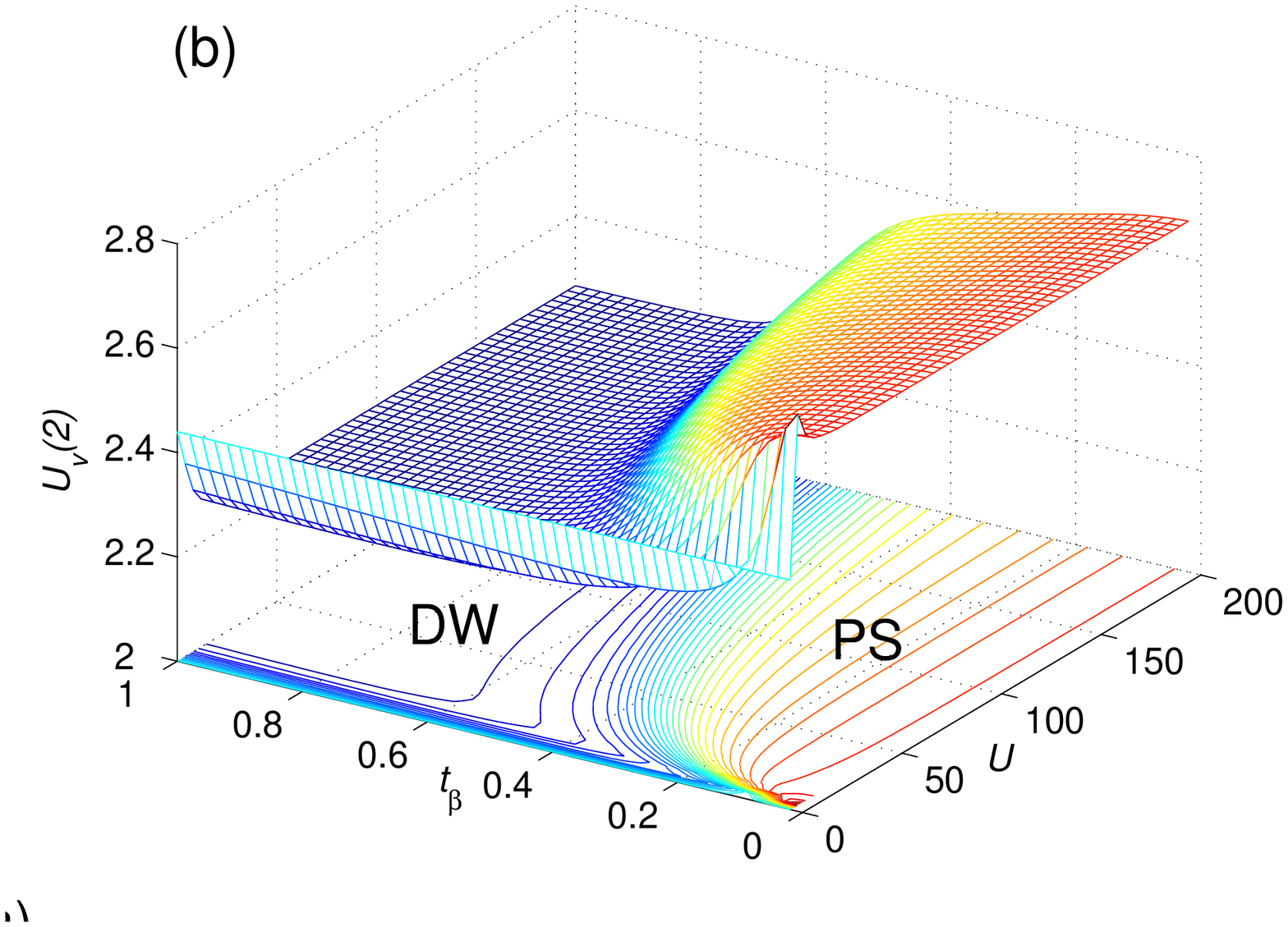}\\
\includegraphics[width=7.5cm]{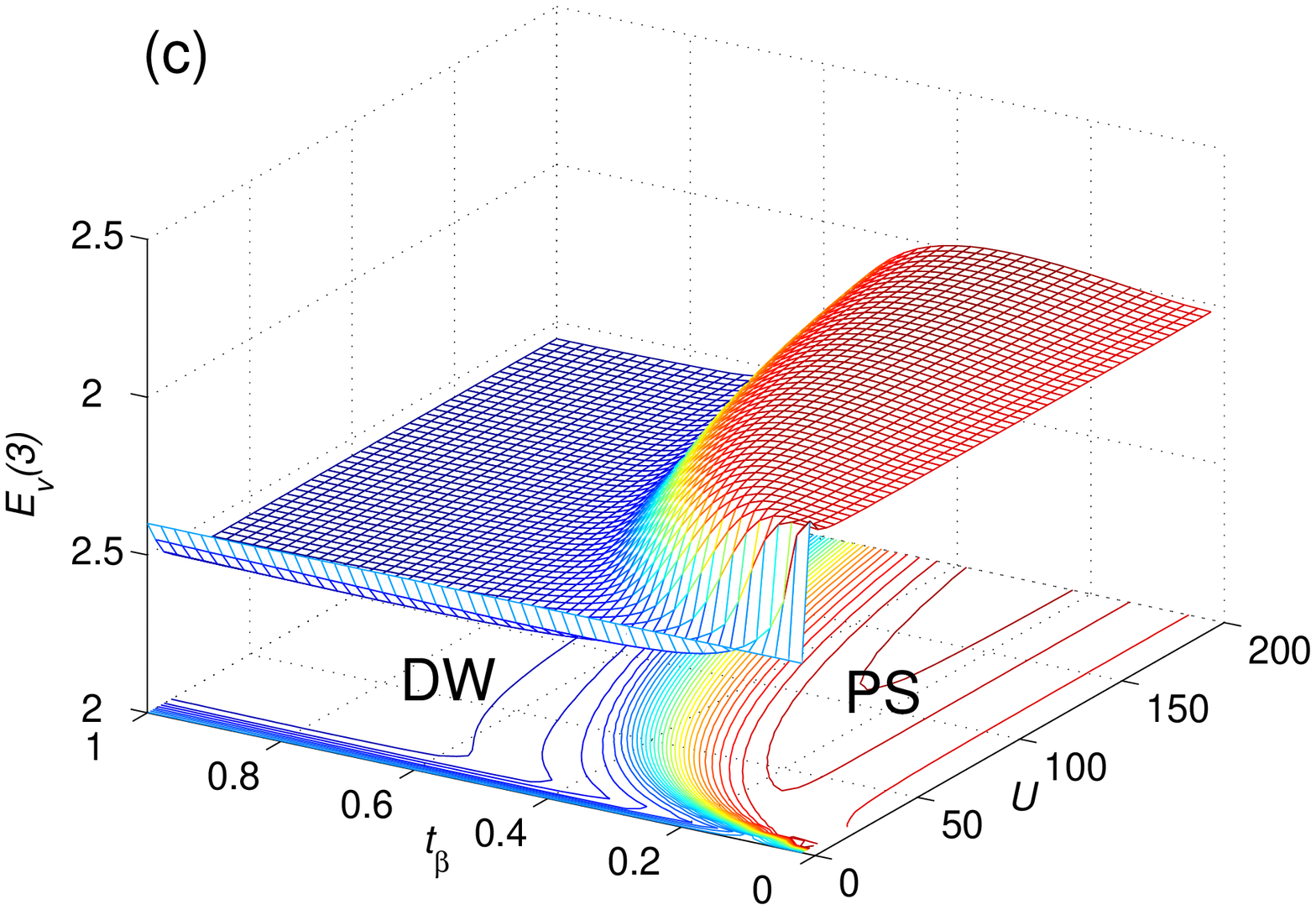}
\includegraphics[width=7.5cm]{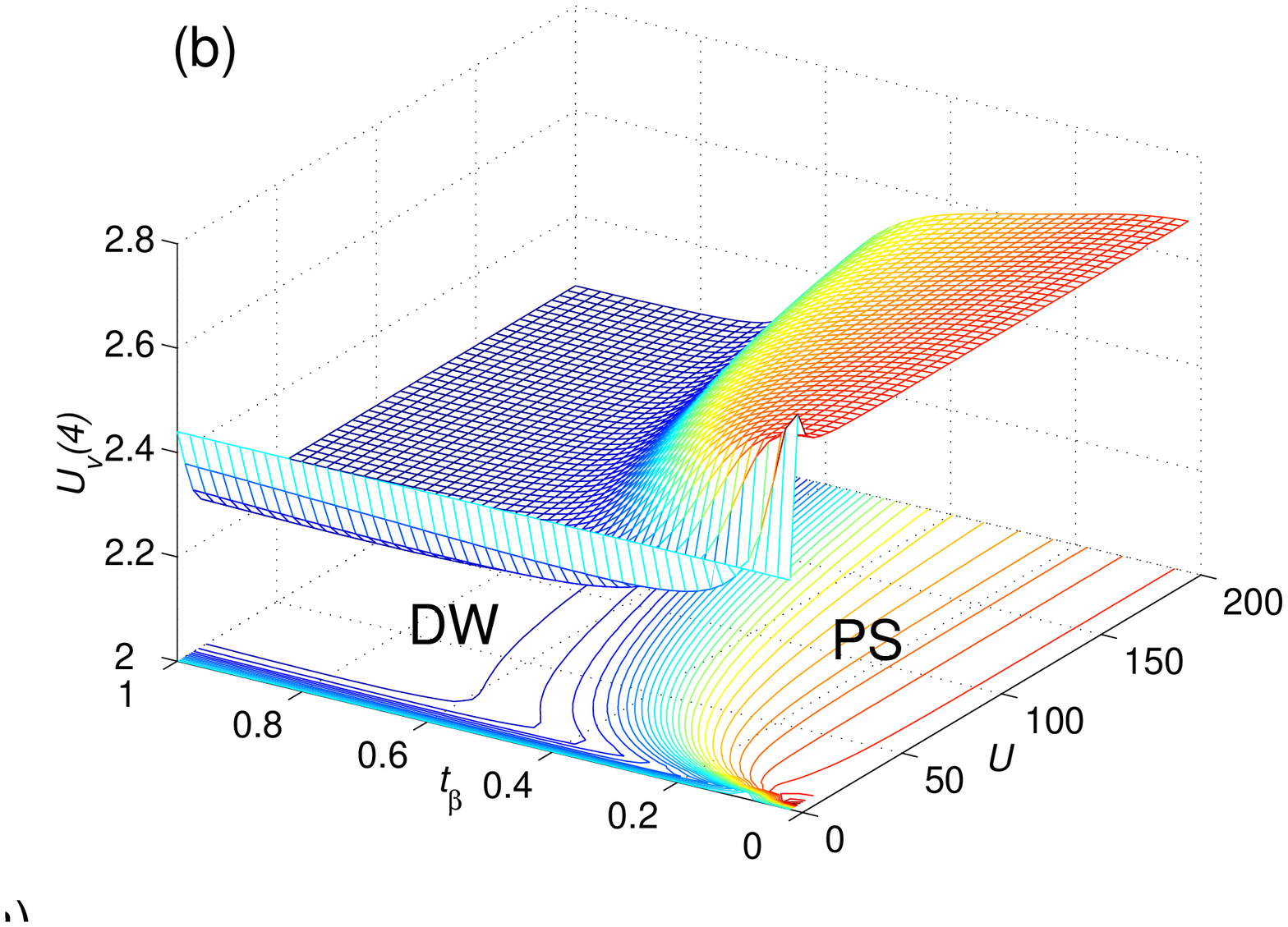}
\caption{(color online) The changes of symmetry in the ground state
wave function is analyzed by considering the quantum correlation,
i.e. entanglement, between local block and other parts of the
system. Here $L=6, N_\alpha=N_\beta=2, l=2$, and the anti-periodic
boundary conditions are assumed in order to avoid level-crossing in
the ground state. Four figures correspond
to different block size: a($l=1$), b($l=2$), c($l=3$), and d($l=4$).\\
\label{fig:6lent2}}
\end{figure*}

\begin{figure}
\includegraphics[width=7.5cm]{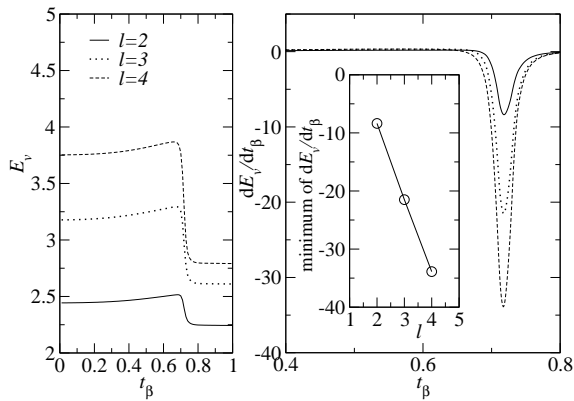}
\caption{The entanglement (left) and its first derivative (right) as a function
of $t_\beta$ for various block size and a specified $U=200$.  The inset shows
the scaling behavior of the minimum point of $dE_v(l)/dt_\beta$ at the critical
point. Here $L=10, N_\alpha=N_\beta=4$.\\\label{fig:lent_200} }
\end{figure}

\section{Ground state entanglement}
\label{sec:ent}

In recent years, studies on the role of entanglement in the quantum critical
behavior \cite{Sachdev} have established a bridge between quantum information
theory and statistical physics. \cite{QEQPTExample,GVidal2003,SJGuPRL} It is
believed that the entanglement, as a kind of quantum correlation, can help us
identify quantum phase transition in many-body systems. To have an intuitive
picture of the global phase diagram, we must first compute the entanglement
between a local block and the rest of the system. For the present model, the
local states on each site have four possible configurations, denoted by
$$\phi_l=|0\rangle,\,|\alpha\rangle,\,|\beta\rangle,\,
|\alpha\beta\rangle;\,l=1,2,3,4.$$ The Hilbert space associated with the
$L$-site system is spanned by $4^L$ basis vectors. If we choose the periodic
boundary conditions for $N=4n+2$ and antiperiodic boundary conditions for
$N=4n$, where $n$ is an integer, the ground state is nondegenerate. Considering
the reduced density matrix of a block of $l$ successive sites of the ground
state
\begin{eqnarray}
\rho_l={\rm tr}_r |\Psi\rangle \langle\Psi|,
\end{eqnarray}
the von Neumann entropy $E_v(l)$, i.e.
\begin{eqnarray}
E_v(l)=-{\rm tr}[\rho_l\log_2(\rho_l)]
\end{eqnarray}
measures the entanglement between the $l$ sites and the $L-l$ sites of the
system. Like the well known fact in classical optics that the three-dimensional
image of one object can be recovered from a small piece of holograph due to the
interference pattern of the reflected light beams from it, quantum
superposition principle also allows us to see a global picture of the system
from its local part \cite{GVidal2003,SJGuPRL}. As was shown for some typical
models in condensed matter physics, such as the extended Hubbard
model\cite{SJGuPRL}, the entanglement of the ground state can give us a global
view of the phase diagram.

>From this point of view, we show a three-dimensional diagram and its contour
map of the entanglement with block size $l=1, 2, 3, 4$ for 6-site system with
$N_\alpha=N_\beta=2$ in Fig. \ref{fig:6lent2}. It has been pointed out that in
the extended Hubbard model, the single-site entanglement can distinguish the
three main phases in the ground state. The reason is that the density
distributions of the different modes in the reduced density matrix of a single
site, such as the double occupancy, in the extended Hubbard model are sensitive
in the quantum phase transitions. However, from Fig. \ref{fig:6lent2}(a), the
single-site entanglement in AHM is rather trivial in the large $U$ region. It
is not difficult to understand this phenomena. For the present model, the
reduced density matrix of a single site has a simple form,
\cite{PZanardi_PRA_65_042101,SJGuPRL},
\begin{eqnarray}
\rho_1= z|0\rangle\langle 0| + u^+|\alpha\rangle\langle\alpha|
         + u^- |\beta\rangle\langle\beta|
          + w |\alpha\beta\rangle\langle\alpha\beta| ,
\end{eqnarray}
in which $z, u^+, u^-$, and $w$ are the density distributions for different
local states, and can be calculated as
\begin{eqnarray}
w&=&\langle n_{\alpha}n_{\beta}\rangle
 = {\rm tr}(n_{ \alpha}n_{ \beta}\rho_1), \nonumber\\
  u^+&=&\langle n_\alpha\rangle - w,  \;\;
   u^-=\langle n_\beta\rangle - w, \nonumber\\
    z&=&1 - u^+ - u^- -w.
\end{eqnarray}
In the large $U$ limit, the double occupancy of two atoms on a single site is
forbidden, i.e. $w\simeq 0$. Then for a finite system with periodic or
anti-periodic boundary conditions, $\langle n_\alpha\rangle$ and $\langle
n_\beta\rangle$ are constants. This fact leads to a constant single-site
entanglement during the evolution of $t_\beta$ in the large $U$ region (Fig.
\ref{fig:6lent2}(a)). For the case of $n_\alpha=1/3, n_\beta=1/3$, $E_v(1)$ has
the value $\log_2 3$, so the single-site entanglement is insensitive to the
phase transition from DW to PS. This property is very similar to the
single-site entanglement in spin models \cite{SJGUNJP} and the ionic Hubbard
model.\cite{Legeza06}

Clearly, the transition from DW to PS is intrinsically related to the change of
density distribution of one species of atoms on the lattice. In order to
contain enough information of the density-density correlation from the point of
view of the entanglement, more sites should be included into the block.
According to this point, we show the two-site entanglement as a function of
$t_\beta$ and $U$ in Fig. \ref{fig:6lent2} (b), from which we immediately
notice two different regions: one is an altiplano marked with warm color
(denoted by ``PS'' in the contour map of Fig. \ref{fig:6lent2} (b)), while the
other is a plain with cold color (denoted by ``DW'' in contour map of Fig.
\ref{fig:6lent2} (b). Taking into account the known fact of the two limiting
cases of this model, such an obvious difference witnesses the critical
phenomenon between two universal classes.

In order to understand this obvious difference of the two-site entanglement in
two phases, let us have a look at the structure of the corresponding reduced
density matrix. For the AHM, the total numbers of $\alpha$ atoms and $\beta$
atoms are good quantum numbers, which leads to the fact that for arbitrary
block size $l$, there is no coherent superposition of local states with
different values of $N_\alpha$ and $N_\beta$ in the reduced density matrix.
That is, the reduced density matrix must have the block-diagonal form
classified by both $N_\alpha(l)$ and $N_\beta(l)$, i.e.
\begin{eqnarray}
\rho_l={\rm diag}\{\rho_l(0,0), \rho_l(1, 0), \rho_l(0,1), \dots, \rho_l(l, l)\}
\end{eqnarray}
where $\rho_l(n_a, n_b)$ is a matrix which has $n_a$ $\alpha$ atoms and $n_b$
$\beta$ atoms. According to the definition of von Neumann entropy (hence the
entanglement), its magnitude is really determined by the distribution of the
eigenvalues of the reduced density matrix. That is, the more uniformly
distributed the eigenvalues, the higher the entropy. In the PS phase, elements
in the reduced density matrix related to the basis $|\beta\beta\rangle$, which
denotes two $\beta$ atoms congregate together, are finite, while in the DW
region, they are almost zero. This fact leads to a larger entanglement with a
block size larger than 2 in the PS phase, but otherwise in the DW phase. So the
transition introduces a significant change into the value of the entanglement,
and vice versa. From Fig. \ref{fig:6lent2} (c), we can see that $E_v(3)$ shares
similar properties with $E_v(2)$. On the other hand, since the ground state is
translational invariant, the entanglement satisfies the equation
$E_v(l)=E_v(L-l)$, Therefore, we have the same figures of $E_v(2)$ and $E_v(4)$
(Fig. \ref{fig:6lent2} (d)) for the 6-site system.

Moreover, in the region $l\in[0, L/2]$ the entanglement is a non-decreasing
function of $l$, as is shown in Fig. \ref{fig:lent_200} for a 10-site system
with $N_\alpha=N_\beta=4$ and $U=200$. Therefore its first derivative develops
a minimum at the critical point, as we can see from Fig. \ref{fig:lent_200}.
Moreover, as the block size increases, the minimum point becomes sharper and
sharper, exhibits a scaling behavior as shown in the inset of Fig.
\ref{fig:lent_200}, i.e., $dE_v(l)/dt_\beta \propto -l$ around the critical
$t_\beta$.

\begin{figure}
\includegraphics[width=7.5cm]{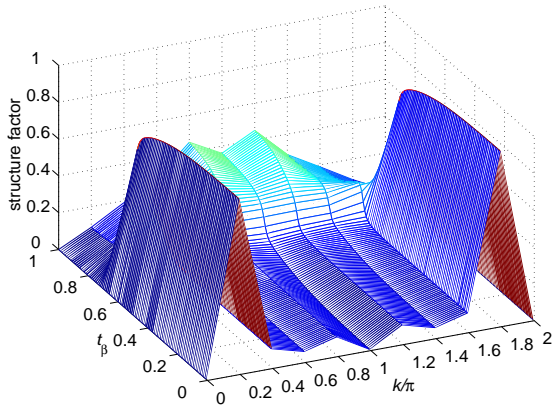}
\caption{(color online) The structure factor of DW as a function of $t_\beta$
and various modes, i.e. quantized momentum. Here $L=10, N_\alpha=N_\beta=4,
U=200$.\\ \label{fig:cdwc200} }
\end{figure}

\begin{figure}
\includegraphics[width=7.5cm]{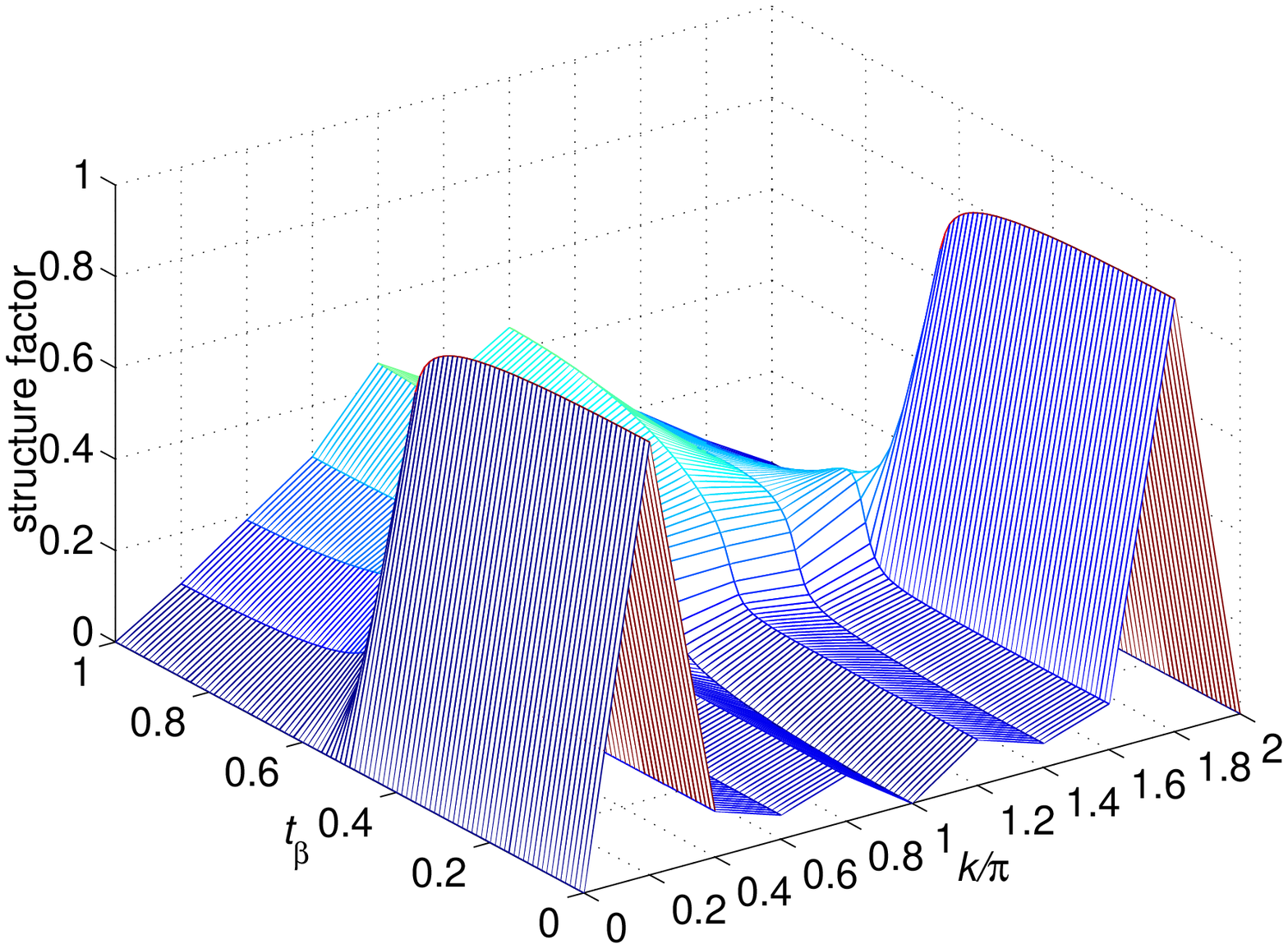}
\caption{(color online) The structure factor of DW as a function of $t_\beta$
and various modes, i.e. quantized momentum. Here $L=10, N_\alpha=N_\beta=4,
U=50$.\\ \label{fig:cdwc50} }
\end{figure}

\begin{figure}
\includegraphics[width=7.5cm]{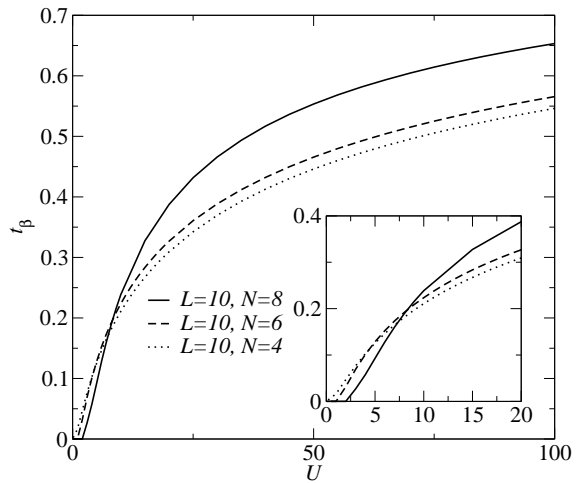}
\caption{(color online) The ground phase diagram in the $U-t_\beta$ plane in
large $U$ case. The dominating configuration in the up-region is DW and below
the boundary line it is PS. Here $L=10, N_\alpha=N_\beta$.\\\label{fig:phase} }
\end{figure}

\section{Charge order parameter and phase diagram}
\label{sec:cdwo}

Though the entanglement can give us useful information about the phase diagram,
the dominating configurations in different phases remains unknown. So it is
important to study the structure factor in competing phases. Taking into
account that the dominating configuration of $\beta$ atoms is quite different
in two phases, we introduce the following structure factor of DW of $\beta$
atoms
\begin{eqnarray}
S_{\rm CDW}(q) =\frac{1}{L}\sum_{jl}\left[ e^{iq(j-l)}(\langle n_{j,\beta}
n_{l,\beta} \rangle - \langle n_{\beta}\rangle^2) \right],
\end{eqnarray}
where $q=2n\pi/L,\, n=0, 1, \cdots, L$. In  Fig. \ref{fig:cdwc200}, we show
that the structure factor as a function of $t_\beta$ for different modes for a
system with $L=10$ and a relatively large $U=200$. The figure shows an obvious
competition between the two modes. In the small $t_\beta$ limit, i.e., when
$\beta$ atom has very heavy mass, $S_{\rm CDW}(q=2\pi/L)$ dominates, which
indicates phase separation in this region \cite{PLemberger92}. A careful
scrutiny of the ground-state wavefunction finds that the configuration
$$|\beta, \beta, \beta,\beta, \circ, \circ, \circ, \circ, \circ, \circ\rangle
$$ of $\beta$ atoms is dominant. It is not difficult to interpret
this result. In the small $t_\beta$ limit, the major contribution to the
ground-state energy comes from the $\alpha$ atoms. So in order to have a lower
energy, they need more free space because the energy of particles inside the
Fermi surface is $-2\cos(k), k\propto 1/L$. Then $\beta$ atoms will be pushed
by $\alpha$ atoms to form clusters and phase separation occurs. For a finite
size system, the translational symmetry is still preserved, while in the
thermodynamic limit, especially in high dimensions, the symmetry in the ground
state might be broken due to the very high potential energy between different
configurations of $\beta$ atoms. Then the system will be separated into two
distinct regions macroscopically. One is the solid-like region of $\beta$
atoms, while the other is the liquid-like region of $\alpha$ atoms. The latter
can be described by a model of $N_\alpha$ atoms trapped in an infinite
potential well with length $L-N_\beta+1$. The ground state is insulating and
the energy is simply
\begin{eqnarray}
E_0\simeq -2\sum_{j=1}^{N_\alpha} \cos\left(\frac{j\pi}{L-N_\beta+1}\right).
\end{eqnarray}
While if $t_\beta\rightarrow 1$, $S_{\rm CDW}(q=N\pi/L)$ exceeds $S_{\rm
CDW}(q=2\pi/L)$, which implies that $\beta$ atoms distribute uniformly on the
optical lattice. Then together with $\alpha$ atoms, the ground state becomes
the so called DW state, which can be regarded as a solution of $\alpha$ and
$\beta$ atoms, as shown by the configuration $$|\beta, \alpha, \circ, \beta,
\alpha, \beta, \circ, \alpha, \beta, \alpha \rangle.
$$ of $\alpha$ and $\beta$ atoms. In its limiting case $t_\beta=1$, the model
goes back to the traditional Hubbard model whose excitation spectrum is
gapless, so the system is a conductor away from half-filling\cite{Hubbard}.
Therefore, different configurations dominate in different regions and the
competition between them leads to a critical phenomenon. According to this
criterion, we can use the intersection of the structure factor of two modes to
determine the transition point on the  $U-t_\beta$ plane for a finite system.
We plot the phase diagram on the $U-t_\beta$ plane in Fig. \ref{fig:phase} for
a 10-site system with different filling $N_\alpha=N_\beta=4, 6, 8$.

\begin{figure}
\includegraphics[width=7.5cm]{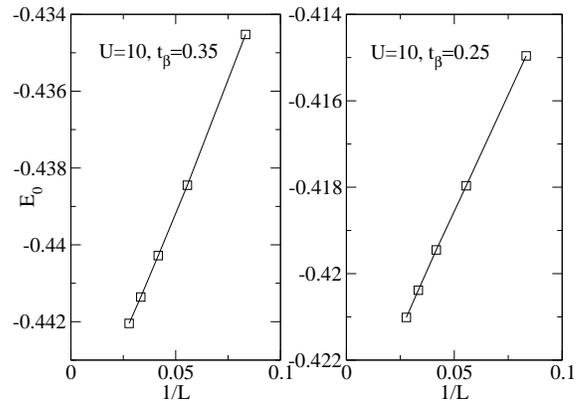}
\caption{The scaling behavior of the ground-state energy for two points near to
the both sides of the critical point. Here $n=2/3$.\\\label{fig:scale_ge} }
\end{figure}

\begin{figure}
\includegraphics[width=7.5cm]{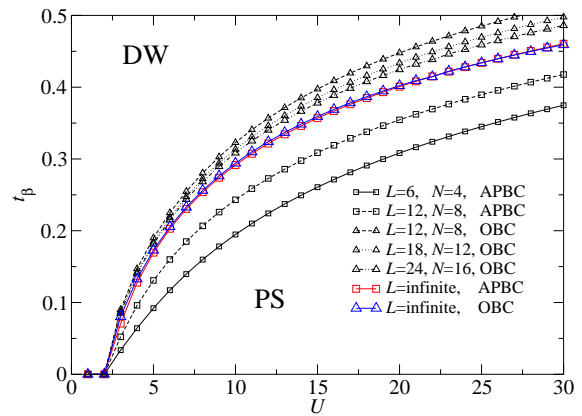}
\caption{The scaling behavior of the phase boundary for both open boundary
conditions (triangular lines) and anti-periodic boundary conditions (square
lines). Here $n=2/3$.\\\label{fig:scale_phaseboundary} }
\end{figure}

However, the results for a finite system is rather qualitative. In order to
have quantitative results for a real system, scaling analysis is crucial. For
this purpose, we first estimate the scaling behavior of the ground-state energy
in the critical region by the ED and DMRG method.\cite{DMRG} In Fig.
\ref{fig:scale_ge}, we show the scaling behavior of the ground-state energy at
a given density $n=2/3$. Results are obtained for systems with open boundary
conditions via the DMRG method in which up to 150 states are kept in the finite
algorithm. It is evident that the limiting energy is approached linearly with
$1/L$. A relation of the form
\begin{eqnarray}
E_0(N)=E(\infty)+a/L, \label{eq:limitingenergy}
\end{eqnarray}
where $a$ is a constant, holds quite accurately in a large $U$ region. Since
the quantum critical phenomena is related to the singularity in ground-state
energy, the $1/L$ correlation in Eq. (\ref{eq:limitingenergy}) actually implies
that the phase boundary bears a similar scaling behavior. Based on this
consideration, we take $n=2/3$ as an example to show the scaling behavior of
the phase boundary for both open and anti-periodic boundary conditions in Fig.
\ref{fig:scale_phaseboundary}.

\begin{figure}
\includegraphics[width=7.5cm]{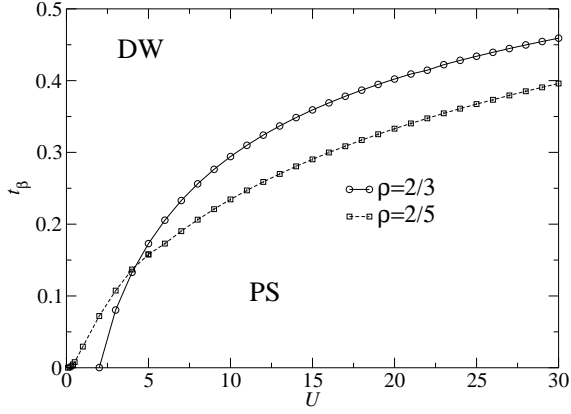}
\caption{The ground-state phase diagram of the AHM for two filling condtions:
$n=2/3, 2/5$.\\\label{fig:phaseboundary} }
\end{figure}

It has been shown that, based on the variation principles, one can obtain the
lower and upper bounds of the phase boundary with different boundary
conditions, such as periodic, anti-periodic, and open boundaries. From Fig.
\ref{fig:scale_phaseboundary},  it is clearly shown that data with APBC give a
lower bound while data with OBC give an upper bound on the transition point.
Moreover, the extrapolated data based on the $1/L$ scaling of the two
approaches for an infinite system agree with each other. This phenomenon is
consistent with the fact that the physics in a real system should be
independent of the boundary conditions. Moreover, we can also estimate errors
in our extrapolation. We presented a final phase diagram with error bars
smaller than the size of the symbols in Fig. \ref{fig:phaseboundary}.

In the small $U$ region, we can see that there is a critical $U$ on the
$U$-axis for the density $n=2/3$. However, if $n$ is reduced, the critical $U$
tends to zero. In the low density limit, the phase boundary scales like
$t_\beta\propto U^2$ in the small $U$ region, which agrees with the results
obtained by the Bosonization method \cite{ZGWangb} excellently.

\begin{figure}
\includegraphics[width=7.5cm]{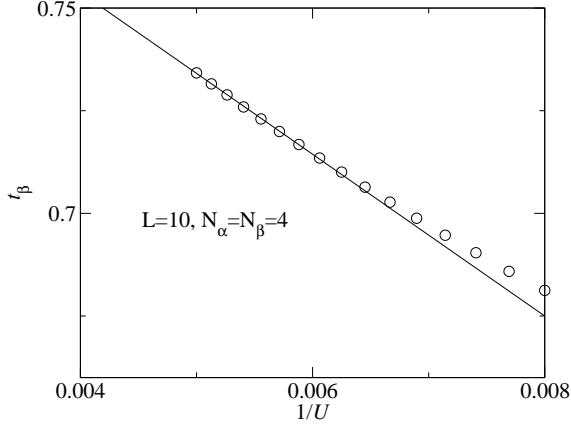}
\caption{$1/U$ behavior of the phase boundary in the large $U$ region, as
exemplified by a system with $L=12, N_\alpha=N_\beta=4$.
\\\label{fig:phase12_8_log} }
\end{figure}

In the large $U$ region, the critical $t_\beta$ increases as $U$ increases. We
take a system of $L=12, N_\alpha=N_\beta=4$ as an example, and show the $1/U$
behavior of the phase boundary in Fig. \ref{fig:phase12_8_log}. From the
figure, we can see that the critical $t_\beta$ is proportional to $1/U$ in the
large $U$ limit. Moreover, Fig. \ref{fig:phase12_8_log} manifests that $U$ will
be saturated in the infinite $U$ limit. That is, for a given concentration,
there exists a saturation $t_\beta^s$ above which the phase separation will
never happen regardless how large the on-site $U$ is. Based on these physical
intuition, the boundary line satisfies the relation
\begin{eqnarray}
t_\beta = t_\beta^s+ C/U, \label{eq:criticaltlu}
\end{eqnarray}
where $C$ is a constant, and both $t_\beta^s$ and $C$ depend on the filling
conditions.

\section{Single-hole problem}
\label{sec:onehole}

In this section, we give a rigorous proof that even for the case of one hole
doping away from half-filling, the DW state is unstable to the PS state in the
infinite $U$ limit. If $U$ is very large, the critical point then is approached
linearly with $1/U$.

\begin{figure}
\includegraphics[width=7.5cm]{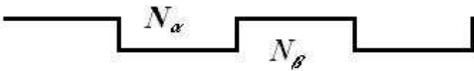}
\caption{The periodical square potential wells used to describe the
dynamics of a hole in demixed phase.\\\label{fig:ppen}}
\end{figure}

We first consider an odd-site sample with $L=2N_\alpha+1, N_\alpha=N_\beta$ and
infinite $U$. The space of DW is spanned by $2L$ basis: $$|e_i\rangle =
|\alpha_1,\beta_2,\dots\circ_i\dots,\alpha_{L-1},\beta_L\rangle, i\in[1, L]$$
and $$|e_{i}\rangle
=|\beta_1,\alpha_2,\dots\circ_{i-L}\dots,\beta_{L-1},\alpha_{L}\rangle,
i\in[L+1, 2L],$$ where $\circ_i$ denotes a hole at site $i$. Then the
Hamiltonian (\ref{eq:Hamiltonian}) becomes
\begin{eqnarray}
\left(%
\begin{array}{cccccccc}
  0 & -1 & 0 & 0 & \dots & 0& -t_\beta \\
 -1 & 0 & -t_\beta & 0& \dots  & 0& 0 \\
  0 & -t_\beta & 0 & -1& \dots & 0 & 0 \\
  0 & 0 & -1 & 0 & \dots & 0 & 0\\
\vdots & \vdots & \vdots & \vdots & \ddots & \vdots& \vdots\\
 0 & 0 & 0 & 0 & \dots & 0 & -1 \\
-t_\beta &0 &0 & 0 & \dots & -1 & 0 \\
\end{array}%
\right).
\end{eqnarray}
It can be solved exactly. The whole energy spectra of the system are given by
\begin{eqnarray}
E_{\pm}=\pm\sqrt{1+t_\beta^2+2t_\beta\cos(k_j)},\\ \nonumber k_j=j\pi/L, \,
j=0,1,\dots,L-1
\end{eqnarray}
The ground-state wavefunction for arbitrary odd $L$ is
\begin{eqnarray}
|\Psi_0\rangle=\frac{1}{\sqrt{2L}}\sum_{i=1}^{2L} |e_i\rangle,
\end{eqnarray}
with the eigenenergy $E_{\rm DW}=-1-t_\beta$.

However, in the case of completely demixed phase, the whole space is spanned by
$L(L-1)$ basis. The Hamiltonian then describes the problem of single particle
motion in periodic square potential wells (see Fig. \ref{fig:ppen}) with
different hoping integrals in different region. Precisely, the Hamiltonian
becomes
\begin{eqnarray}
H=-\sum_{j,\delta}d_j^\dagger d_{j+\delta},
\end{eqnarray}
where $d_j^\dagger$ and $d_j$ are hole creation and annihilation operators in
each potential well, and
\begin{eqnarray}
H=-t_\beta\sum_{j,\delta}d_j^\dagger d_{j+\delta}
\end{eqnarray}
elsewhere. The ground state of this Hamiltonian in the thermodynamic limit is
identical to the ground state of
\begin{eqnarray}
H=\left\{ \begin{array}{cc}
  -2+p^2, & 0<x<N_\alpha; \\
  -2t_\beta+ t_\beta p^2, & N_\alpha<x< N_\beta, \\
\end{array}
\right.
\end{eqnarray}
with periodic boundary conditions
\begin{eqnarray}
\Psi(x)=\Psi(x+N_\alpha+N_\beta).
\end{eqnarray}
For the latter, a bound state always exists for arbitrary well depth
$2-2t_\beta$. Then if $N_\alpha=N_\beta\rightarrow\infty$, the ground-state
energy is simply -2, which is obviously smaller than $E_{\rm DW}=-1-t_\beta$.
For a finite system except $L=3$, it can also be shown that $E_{\rm PS}(L) <
E_{\rm DW}(L)$. For example if $L=5, N_\alpha=N_\beta=2$, we have
\begin{eqnarray}
E_{\rm PS}(L=5)=-\sqrt{2+2t_\beta^2}.
\end{eqnarray}
Therefore, in the infinite $U$ limit, the DW state is unstable to the PS state.
Such a rigorous result is also valid for a system of two species of hard-core
bosonic atoms and Bose-Fermi mixtures with different hoping integrals in
optical lattices.

When the on-site $U$ is very large but not infinite, the Hamiltonian
(\ref{eq:Hamiltonian}) can be approximated by \cite{GFath95}
\begin{eqnarray}
H&=&-\sum_{j=1}^{L}\sum_{\delta=\pm 1} \sum_\sigma t_\sigma
c^\dagger_{j,\sigma}c_{j+\delta, \sigma} \nonumber \\ &&+J \sum_{j=1}^L
\left[\mathbf{S}_j\cdot \mathbf{S}_{j+1}+\delta S_j^zS_{j+1}^z - \mu\, n_j
n_{j+1}\right], \label{eq:Hamiltoniantj}
\end{eqnarray}
in which
\begin{eqnarray}
J=\frac{4t_\alpha t_\beta}{U},\;\; \delta=\frac{(t_\alpha -
t_\beta)^2}{2t_\alpha t_\beta},\;\; \mu=\frac{t_\alpha^2 +t_\beta^2}{8t_\alpha
t_\beta}.
\end{eqnarray}
Clearly if $t_\alpha=t_\beta$, the above Hamiltonian becomes the $t-J$ model.
The ground state of the $t-J$ model with a single-hole doping becomes the
Nagaoka ferromagnetism \cite{YNagaoka66} if the contribution from the kinetic
energy in Eq. (\ref{eq:Hamiltoniantj}) exceeds that from the spin-spin
antiferromagnetic interaction. In order to study the condition of PS, we first
suppose that the ground state of the system is phase separated. Then the
ground-state energy can be approximated by
\begin{eqnarray}
E_{\rm PS}\simeq -2+\frac{(L-6)(1+\delta) J}{4},
\end{eqnarray}
On the other hand, if the ground state is in the DW state, the ground-state
energy can be approximated by that of the XXZ chain. For the latter, the
ground-state energy per bond has the form \cite{CNYang66}
\begin{eqnarray}
e_{\rm XXZ}=\frac{1+\delta}{4}-\sinh\phi\left[\frac{1}{2}+2\sum_{n=1}^\infty
\frac{1}{e^{2n\phi}+1}\right]
\end{eqnarray}
where $\cosh \phi=1+\delta$. Then the ground-stat energy of the DW phase
becomes
\begin{eqnarray}
E_{\rm DW}\simeq -1-t_\beta+(L-2) J e_{\rm XXZ}.
\end{eqnarray}
Here, in both $E_{\rm DW}$ and $E_{\rm PS}$, the finite-size correction to the
ground-state energy is not taken into account, so the critical value is
estimated approximately. Despite of this, the qualitative behavior of critical
point is clear, i.e.
\begin{eqnarray}
t_\beta\simeq 1-JL \sinh\phi\left[\frac{1}{2}+2\sum_{n=1}^\infty
\frac{1}{e^{2n\phi}+1}\right] +O(1/U),
\end{eqnarray}
which means only if $U\gg L$, the DW state is unstable to PS state. For a given
$L$, the phase boundary scales like
\begin{eqnarray}
t_\beta=1-C/U, \label{eq:cirtkdkfd}
\end{eqnarray}
in the large $U$ limit. Clearly, Eq. (\ref{eq:cirtkdkfd}) is consistent with
our previous result Eq. (\ref{eq:criticaltlu}).

\section{Discussions}
\label{sec:discuss}

Obviously, unlike the PS in the $t-J$ model \cite{VJEmery90} and the extended
Hubbard model \cite{SJGuPRL}, which is the consequence of attractive
interaction between particles, the PS in the ground state of the AHM is driven
by kinetic energy. Therefore, though our results are based on a one-dimensional
model, the underlying physics is quite general for systems in any dimension.
That is, in the large $U$ limit, the dynamics of a system of two species of
atoms at zero temperature is dominated by the light atoms. In order to have a
lower energy, they need more free space to move. This mechanism forces heavy
atoms to congregate together, so the latter becomes a solid-like object. In
experiment, two separated regions are expected to be witnessed macroscopically.
However, when $t_\beta\rightarrow t_\alpha$, the dynamics of heavy atoms is
comparable to that of light atoms, the exchange interaction drives the system
into a DW state. Therefore, if we consider the PS state as a classical phase
containing solid-like order and the DW state as a quantum region with liquid
properties in the whole system, the transition reported in our work is just an
example of a crossover from the classical region to the quantum region.

Such an interesting transition is expected to be observed in the on-going
experiments on optical lattices. We take a system consisting of two species of
atoms (such as $^6$Li ($\alpha$) and $^{40}$K ($\beta$) with $m_\beta/m_\alpha
\simeq 20/3$ ) as an example. Since the typical scattering length for alkaline
atoms ranges from $40$ to $100\, a_{\rm Bohr}$
\cite{MOlshanii98,Bartenstein05}, and laser wavelength $\lambda=852\, {\rm nm}$
\cite{MGreiner02}. Then from Eq. (\ref{eq:tUrelation4}) we roughly estimate
that PS phase can be observed when $v>0.4$ according the phase diagram in Fig.
\ref{fig:phaseboundary}.

\section{summary}
\label{sec:sum}

In summary, we have investigated the ground-state phase diagram of
two species of fermionic atoms trapped in one-dimensional optical
lattice. By using the ED method, we computed the block-block
entanglement between a local block and rest part for a small system.
We obtained an intuitive picture of phase diagram of the ground
state and found that the entanglement in the PS region is in general
larger than that in the DW region for a finite system. Its first
derivative develops a sharp downward peak and shows scaling behavior
at the critical point. We also analyzed the structure factor of the
DW of $\beta$ atoms by the ED and DMRG method, and found that the
competition between two different configurations in the ground-state
wavefunction leads to a phase transition at the critical point. The
global phase diagram was obtained from the careful scaling analysis
for various-size systems and different boundary conditions.
Therefore, we results firstly gave a quantitative description of the
ground-state phase transition of the AHM away from the half-filling.
Furthermore, we gave a rigorous proof that even for the case of a
single hole doping, the DW state is unstable to the PS in the
infinite $U$ limit. Such a rigorous conclusion clarifies the
physical picture of the phase separation.

\section*{ACKNOWLEDGEMENTS}

We thank Y. G. Chen, Z. G. Wang, X. G. Wen, G. M. Zhang for helpful
discussions, and Miss Daisy Lin for critical reading of the manuscript. H. Q.
Lin thanks Research Center of Quantum Control at Fadan University for its
hospitality where part of this work was carried out. This work is supported in
part by the Earmarked Grant for Research from the Research Grants Council of
HKSAR, China (Project CUHK 401504).

\end{document}